\documentclass[aps,prx,superscriptaddress,showpacs,floatfix,twocolumn]{revtex4}
\usepackage{times}
\usepackage{graphicx}
\usepackage{amsmath}
\usepackage{amssymb}
\usepackage{subfigure}
\usepackage{color}
\usepackage{ulem}
\usepackage{sidecap}
\usepackage{floatrow}

\newcommand{\mH}{\mathcal{H}}
\newcommand{\mT}{\mathcal{T}}
\newcommand{\mV}{\mathcal{V}}
\newcommand{\mP}{\mathcal{P}}
\newcommand{\mS}{\mathcal{S}}

\begin{document}


\title{Edge states and local electronic structure around an adsorbed impurity in a topological superconductor}

\author{Yuan-Yen Tai}
\affiliation{Theoretical Division, Los Alamos National Laboratory, Los Alamos, New Mexico 87545, USA}

\author{Hongchul Choi}
\affiliation{Theoretical Division, Los Alamos National Laboratory, Los Alamos, New Mexico 87545, USA}

\author{Towfiq Ahmed}
\affiliation{Theoretical Division, Los Alamos National Laboratory, Los Alamos, New Mexico 87545, USA}

\author{C. S. Ting}
\affiliation{Texas Center for Superconductivity \& Department of Physics, University of Houston, Houston, Texas 77004, USA}

\author{Jian-Xin Zhu}
\affiliation{Theoretical Division, Los Alamos National Laboratory, Los Alamos, New Mexico 87545, USA}
\affiliation{Center for Integrated Nanotechnologies, Los Alamos National Laboratory, Los Alamos, New Mexico 87545, USA}

\date{\today}

\begin{abstract}
Recently topological superconducting states has attracted a lot of interest. In this work, we consider a topological superconductor with Z$_2$ topological mirror order~\cite{YTai2}  and s$_\pm$-wave superconducting pairing symmetry, within a two-orbital model originally designed for iron-based superconductivity~\cite{YTai1}.
We predict the existence of gapless edge states. We also study the local electronic structure around an adsorbed interstitial magnetic impurity in the system, and find the existence of low-energy in-gap bound states even with a weak spin polarization on the impurity. We also discuss the relevance of our results to the recent STM experiment on Fe(Te,Se) compound with adsorbed Fe impurity~\cite{SHPan}, for which our density functional calculations show the Fe impurity is spin polarized.
\end{abstract}
\pacs{ 73.20.−r, 71.70.Ej, 71.10.Pm, 74.20.Pq}

\maketitle

Topological superconductor (TSC) has been attracted lots of interest on its potential application to the fault-tolerance quantum computation~\cite{Kitaev} and superconducting spintronics~\cite{JLinder}.
The TS can be designed through the interface of a topological insulator with the fully gapped superconductor~\cite{Fu} or by engineering through the interface of ferromagnetic-superconducting nanostructure under several special conditions~\cite{IMartin,MMVazifeh,FPientka,JKlinovaja,IReis,SNPerge}.
Most of these efforts were aimed to look for the {\it edge states} and {\it in-gap bound states} in a fully gapped superconductor.
There are two types of  edge states for a two-dimensional (2D) TSC.
The first type is a {\it chiral} edge state in a time-reversal-symmetry-broken superconducting pairing state like $[p_x+ip_y]_{\uparrow\uparrow}$~\cite{FZhang1,Qi1}. The second type is  a {\it helical} edge state in a time-reversal-symmetry-invariant  pairing state like  $[p_x\pm ip_y]_{\uparrow\downarrow}$)~\cite{Fu,Qi2}.
Recently, the crystalline symmetry protected topological phase with mirror Chern number has been studied in a triplet pairing state of Sr$_2$RuO$_4$~\cite{Ueno}, for which  chiral edge states are obtained~\cite{FZhang1}.
A realization of helical edge state can be established in a junction formed by an $s$-wave superconductor and a strong topological insulator (TI), mimicking a  time-reversal-symmetry-preserved $p$-wave superconducting pairing state~\cite{Fu}.

In this work, we propose a feasible scenario to realize a helical topological mirror superconductor (HTMS) based on the coexistence of the interaction-driven $Z_2$ topological mirror order and $s_\pm$-wave pairing symmetry.
Our concept is based on the nature of a topological metal, where the superconductivity can be generated through a finite Fermi surface with electron-(hole-) Fermi pockets. The $Z_2$ mirror topological order creates robust edge states even in the presence of a fully gapped $s_\pm$-wave pairing symmetry.
We further analyze the local electronic structure of an adsorbed interstitial magnetic impurity  (IMI)
in the 2D bulk of the HTMS, and identify  the different consequence from the topological and non-topological superconductor.
It is important to note that, since the superconductivity is intrinsically inherited from the normal-state band structure, our proposal  avoids the need to  make a TI-SC heterostructure for helical edge states.
To make our model into a real-material context, we  also carry out the DFT calculations for an adsorbed interstitial-Fe impurity in recently discovered 11-family of iron-based superconductors FeSe, and find the adsorbed Fe-atom to be spin polarized. Therefore, our results have a direct relevance to 
the recent STM experiment on Fe(Te,Se) SCs, where a robust zero-energy bound state (ZBS) has been found around an adsorbed Fe atom~\cite{SHPan}.

\begin{figure*}
\includegraphics[scale=0.55]{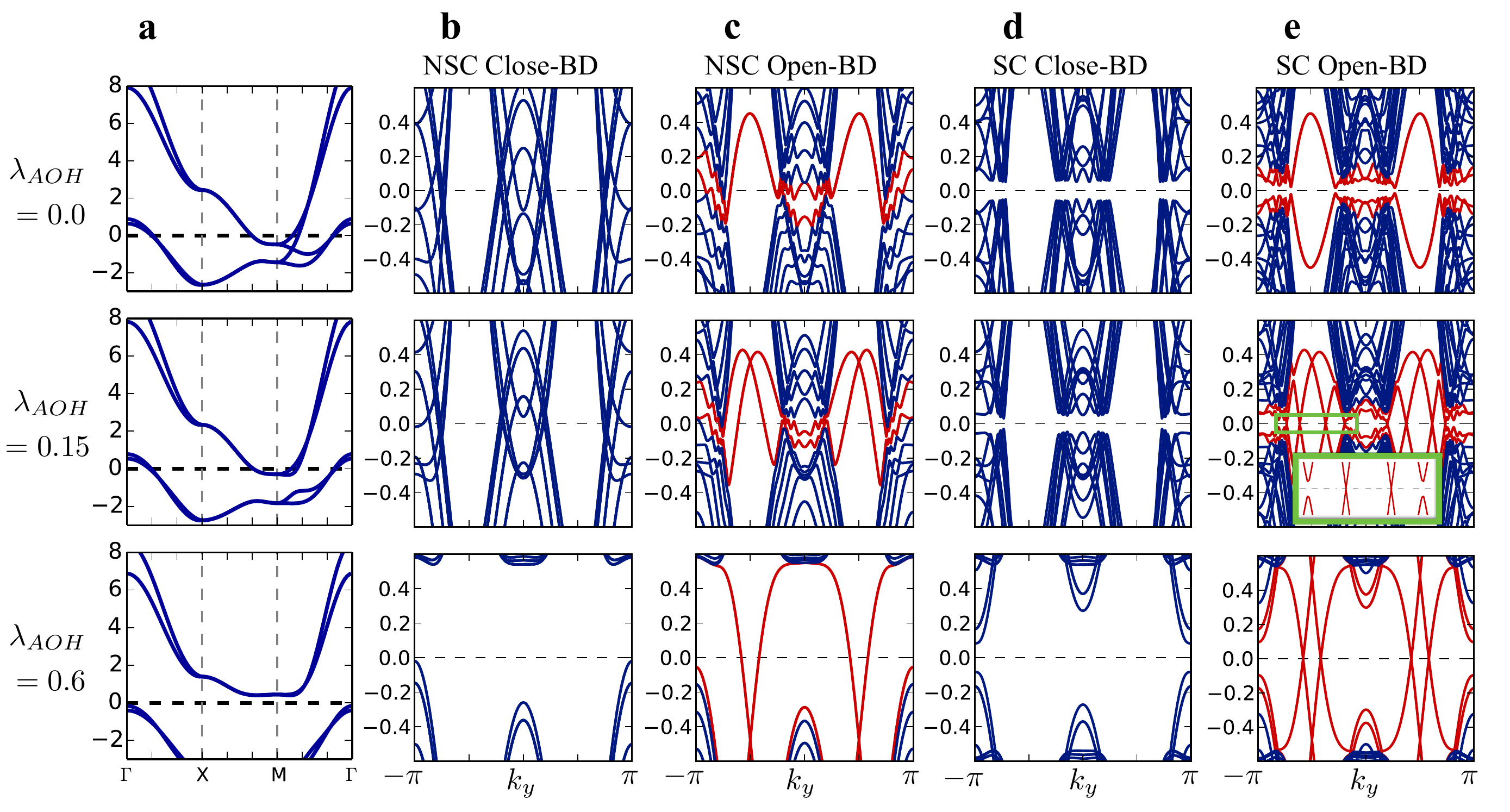}
\caption {
{\bf The band structures shows the gapless edge states of the coexistence of topological and superconducting orders.  }
{\bf a}, The folded band structure of a 2-site per unit cell BZ. {\bf b-e}, The band structure of a strip of width of 20 lattice constants with open and close boundary condition under non-superconducting and superconducting phase. We take the periodicity along the $y$-direction of the strip, where 2000 $k_y$-points are taken. The inset of the middle panel of {\bf e} is the enlargement of the crossing point of the Fermi surface, where we can find that there are totally four degnerated points around $k_y=\pm \frac{\pi}{2}$ (the green inset) and the others around $k_y=0 \text{ and } \pm\pi$ are not degenerated. 
}\label{Bands}
\end{figure*}

We consider the following effective Hamiltonian to describe the coexistence of the $Z_2$ topological mirror order with the $s_\pm$-pairing symmetry,
\begin{equation}
\label{model}
\mH^{TS}=\mT\,[\,t_{1-6},\mu\,]+\mV\,[\,\lambda_{AOH}\,]+\mP\,[\,\Delta\,].
\end{equation}
Here $\mT$ is the kinetic energy with optimized hopping terms ($t_{1-6}$) and chemical potential ($\mu$) to describe the low energy physics that originally designed for the Fe-based compound~\cite{YTai1,HChen}, $\mV$ is responsible for the emergent $Z_2$ topological mirror order~\cite{YTai2} with tunable anomalous orbital Hall order ($\lambda_{AOH}$) for a $d$-wave form factor $\lambda_s({\bf k})=i(-1)^{\alpha\times s}\lambda_{AOH}[\cos(k_x)-\cos(k_y)]_{\alpha,\bar \alpha}$ (where $\alpha\in d_{xz},d_{yz}$, $s\in \uparrow\downarrow$), and
$\mP$ is the $s_\pm$-wave pairing order ($\Delta$) with the form factor $\Delta({\bf k})=4\Delta[\cos(k_x)\times\cos(k_y)]_{\uparrow\downarrow}$ (in 1-Fe BZ).  See definitions of $\mT$, $\mV$ and $\mP$ in the {\bf Method} section. Throughout the work, the chemical potential is adjusted to ensure the system is half filled. In the following section, we numerically solve the band structure of Eq.~(\ref{model})  to reveal the interplay of topological ($\lambda_{AOH}$) and superconducting ($\Delta$) orders. 

\begin{figure}
\includegraphics[scale=0.2]{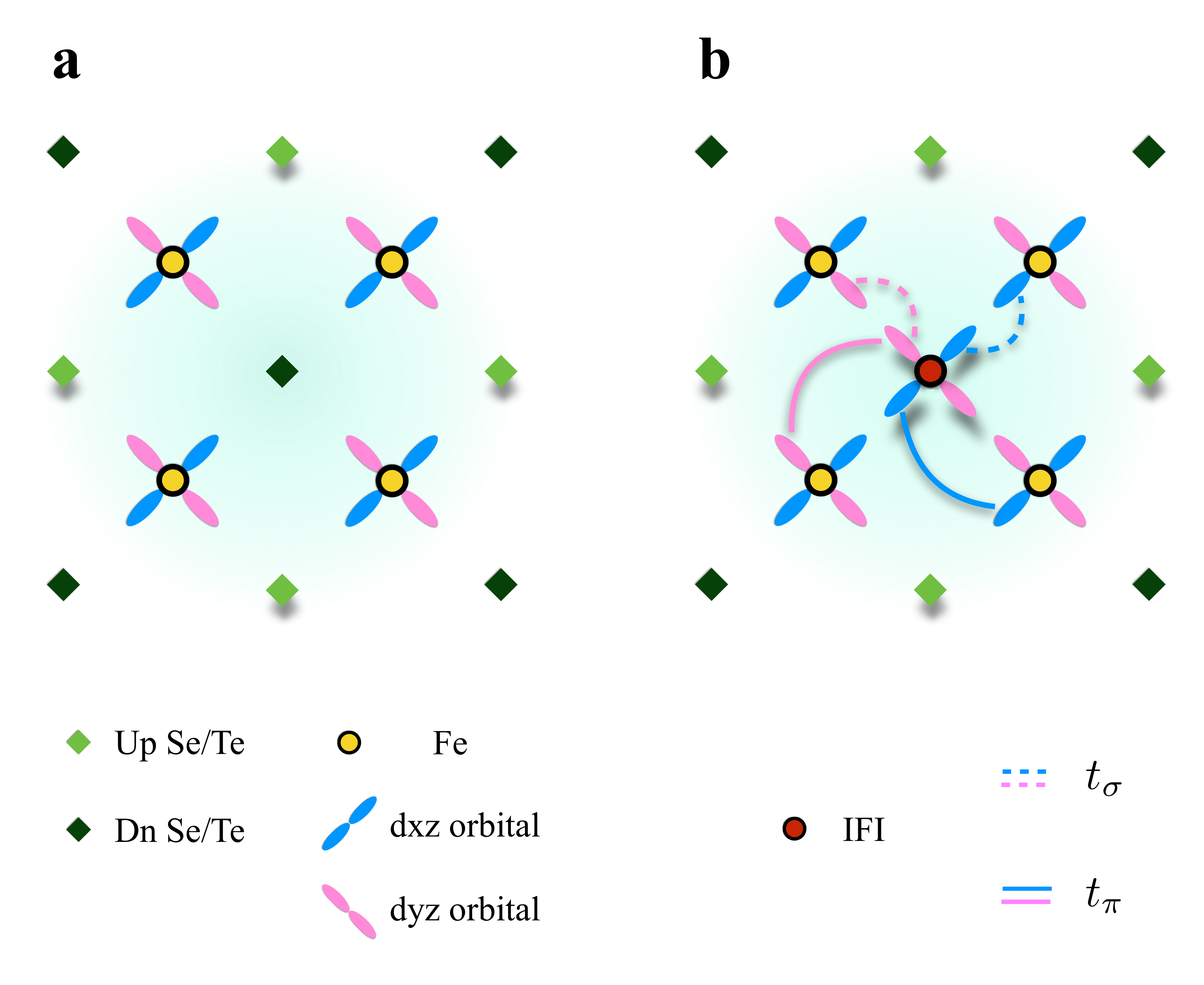}
\caption {
{\bf Cartoon picture for an interstitial impurity in a 2D lattice.}
{\bf a}, The top-view of the parent lattice structure and the orientation of $d_{xz}$ and $d_{yz}$ orbitals.
{\bf b}, The position of an interstitial atom located at the center of a plaquette of the 2D square lattice together with the local hopping process among the $d_{xz}$/$d_{yz}$ orbitals, where $t_{\sigma,\pi}$ stand for orbital lobes aligned parallel (perpendicular) to  the hopping direction between two nearest-neighboring sites of the square lattice).   
We set $t_{\sigma}=a\,t_{\pi}=\xi$ for the parameterization.
}\label{IFI}
\end{figure}

\begin{figure*}
\includegraphics[scale=0.4]{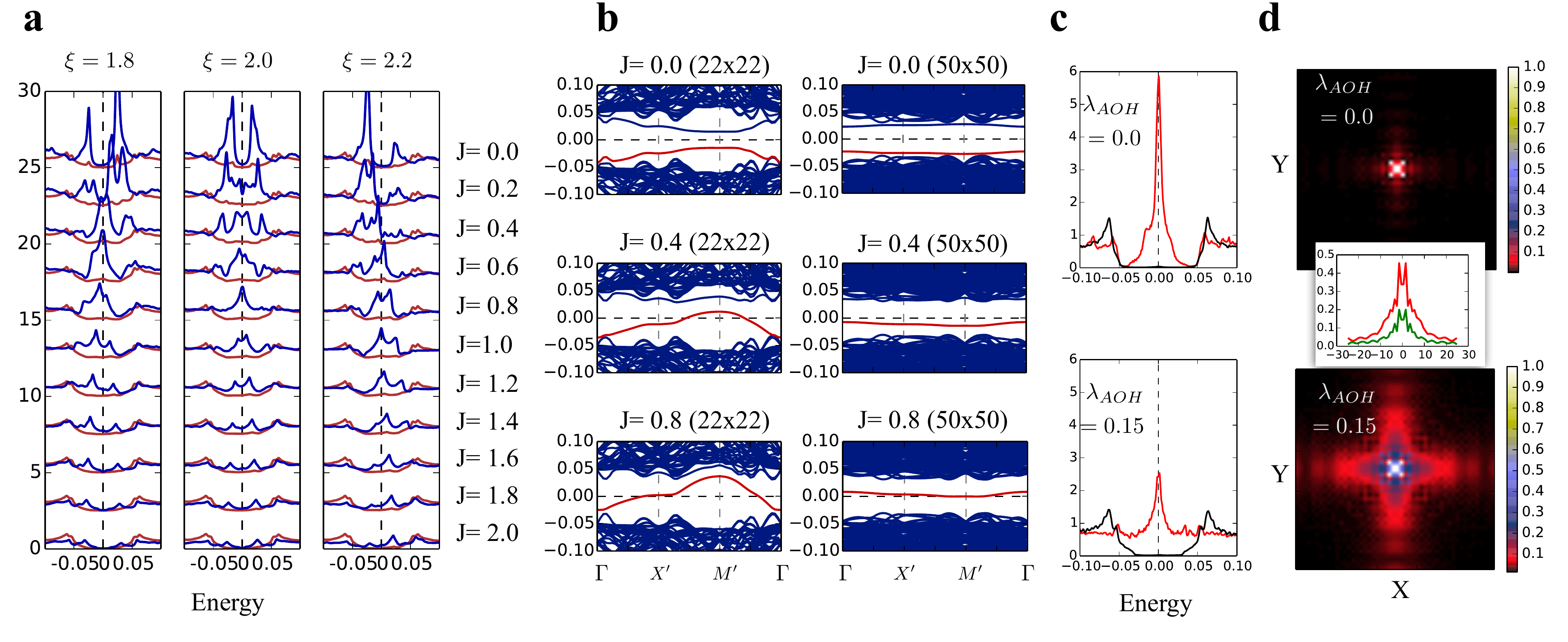}
\caption {
{\bf LDOS and band dispersion from the model Hamiltonian}.
{\bf a}, The calculated LDOS for various values of spin exchange interaction, $J$, on the impurity site (blue lines) and a site deep into the bulk (red lines) as a reference.
{\bf b}, The highly folded band structure for a supercell (of size $N\times N$) lattice  with a single interstitial impurity sitting in the center of each supercell.
{\bf c}, The LDOS on the interstitial impurity site (red-solid lines) and a site deep in the bulk (black-solid lines)   for $\lambda_{AOH}=0.0$ (top panel) and $\lambda_{AOH}=0.15$ (bottom panel) at a fixed $J=0.8$. 
{\bf d}, The spatial dependence of LDOS at zero energy in the entire supercell of the size $50 \times 50$. The inset is the cut from site indeces (-24,0) to (25,0) along a bond direction through the one side of the plaquette, in which the interstitial impurity sits (with site coordinate (0.5,0.5)).  Unless specified otherwise, the order parameter values are $\lambda_{AOH}=0.15$ and $\Delta=0.02$.
}\label{figLDOS}
\end{figure*}

\section{Edge states with fully gapped s$_\pm$-wave pairing}
The TSC property can be revealed by studying the nature of edge states. We  first calculate the energy dispersion under several conditions.
Figure~\ref{Bands}a shows the evolution of the non-superconducting (NSC, $\Delta=0$) band structure with different values of $\lambda_{AOH}$ in Brillouin zone (BZ) corresponding to  the 2-site per unit cell. 
In Fig.~\ref{Bands}a, the band crossing below the Fermi energy occurs  along the $M$-$\Gamma$ direction when there is no existence of the anomalous orbital order (AOH), $\lambda_{AOH}=0$, and electron and hole pockets are formed at the Fermi energy. In the presence of a small value of the AOH term ($\lambda_{AOH}=0.15$),  the band crossing is lifted and the four bands are formed into two disentangled groups. However,  the Fermi pockets remain.
Furthermore, when  the topological order is strong enough (Fig.~\ref{Bands}a with $\lambda_{AOH}=0.6$), an indirect band gap is open directly at the Fermi energy and the bulk system becomes insulating, from which no superconducting pairing can emerge.
Now, we construct a strip geometry to study the electronic states at the edge with periodic (Fig.~\ref{Bands}b,d) and open (Fig.~\ref{Bands}c,e) boundary conditions. Since there still exists the translational invariance along the strip direction (that is, $y$-direction) at the original lattice constant, the momentum along this direction is a good quantum number, which allows to solve the problem for each given wave vector $k_y$ independently. 
 We can see that the edge bands show up by comparing Fig.~\ref{Bands}b,d and c,e. 
In the metallic/topological-metallic states($\lambda_{AOH}=0,\;0.15$) without the superconducting order, the edge bands have curved features across the Fermi surface as highlighted in red in Fig.~\ref{Bands}c. For $\lambda_{AOH}=0$ the non-superconducting (NSC) edge states are four-fold degenerate (including two spin orientations and two edges) at a given $k_y$ near the region of  $k_y=\pm\frac{\pi}{2}$. When $\lambda_{AOH}\neq 0$, the degeneracy of edge bands  near $k_y=\pm\frac{\pi}{2}$ is reduced to two fold (that is, only two-fold  spin degeneracy) at a given $k_y$. However, these bands are still crossing the Fermi energy. We now turn on the $s_{\pm}$-wave pairing ($\Delta=0.02$) (see Fig.~\ref{Bands}d,e). In the absence of the topological order, the edge states have no zero-energy modes, which means the quasiparticle states are always gapped.  Only in the presence of a sufficiently large topological order, can the edge states appear to cross the Fermi energy (see Fig.~\ref{Bands}e with $\lambda_{AOH}= 0.15,\;0.6$). 
We note that for  $\lambda_{AOH}=0.15$, the normal state is metallic and the superconducting pairing is physically feasible. Therefore,  we
 will focus on the case of $\lambda_{AOH}=0.15$ in the next section.
For $\lambda_{AOH}=0.6$ (see third row of Fig.~\ref{Bands}e), the normal state is already insulating and the superconducting pairing is not possible. However, we include this case to show the structure of edge states in response to a ``superconducting'' particle-particle pair. 

\begin{figure}
\includegraphics[scale=0.5]{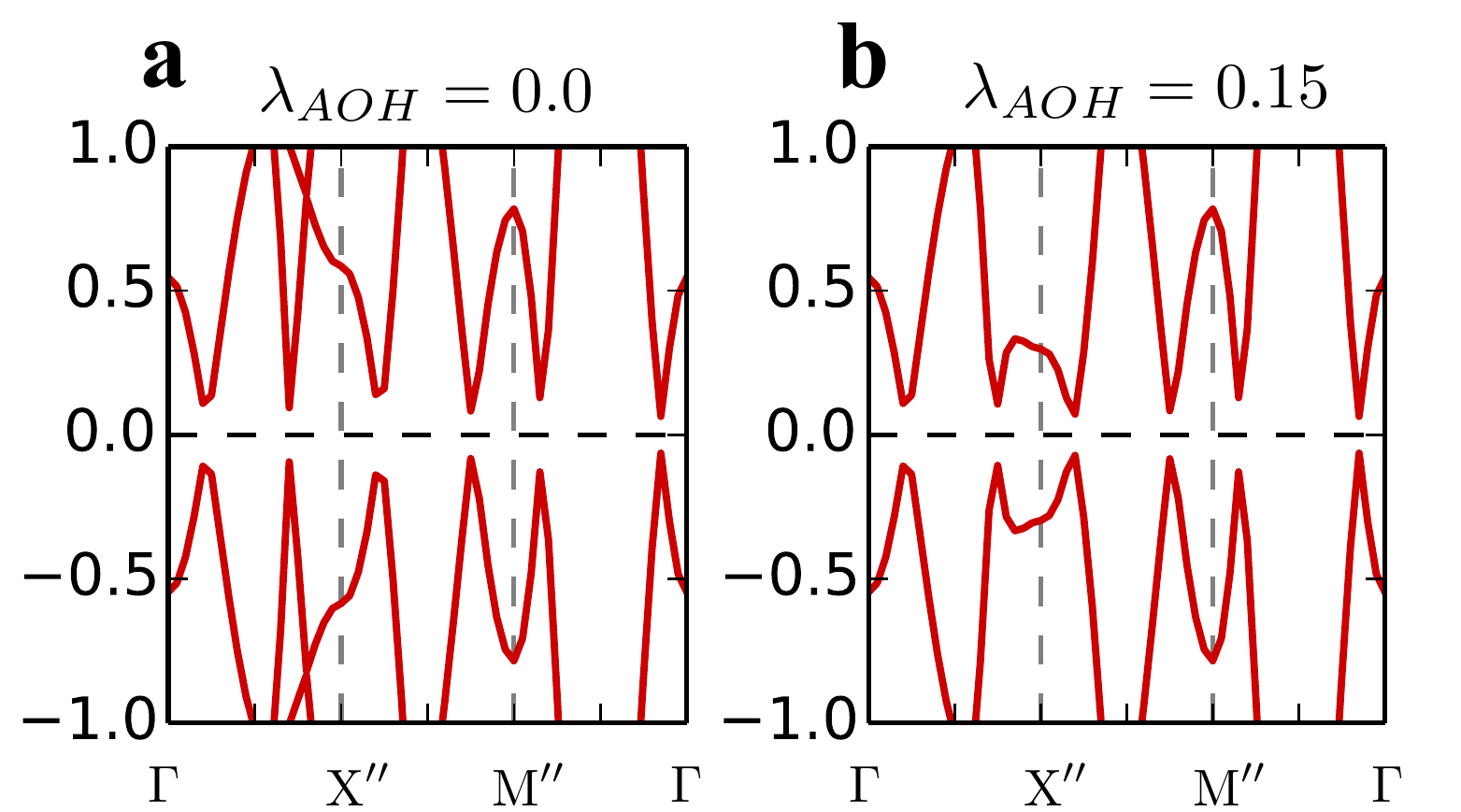}
\caption {
{\bf Band structure from the model Hamiltonian.}
 {\bf a}, $\lambda_{AOH}=0$ and {\bf b}, $\lambda_{AOH}=0.15$.
The high-symmetry momentum points of $\Gamma$, $X^{\prime\prime}$ and $M^{\prime\prime}$ are defined according to the Brillouin zone of a 1-site per unit cell. The superconducting order parameter $\Delta=0.02$ is taken.
}\label{figBdGBand}
\end{figure}

\begin{figure*}
\includegraphics[scale=0.17]{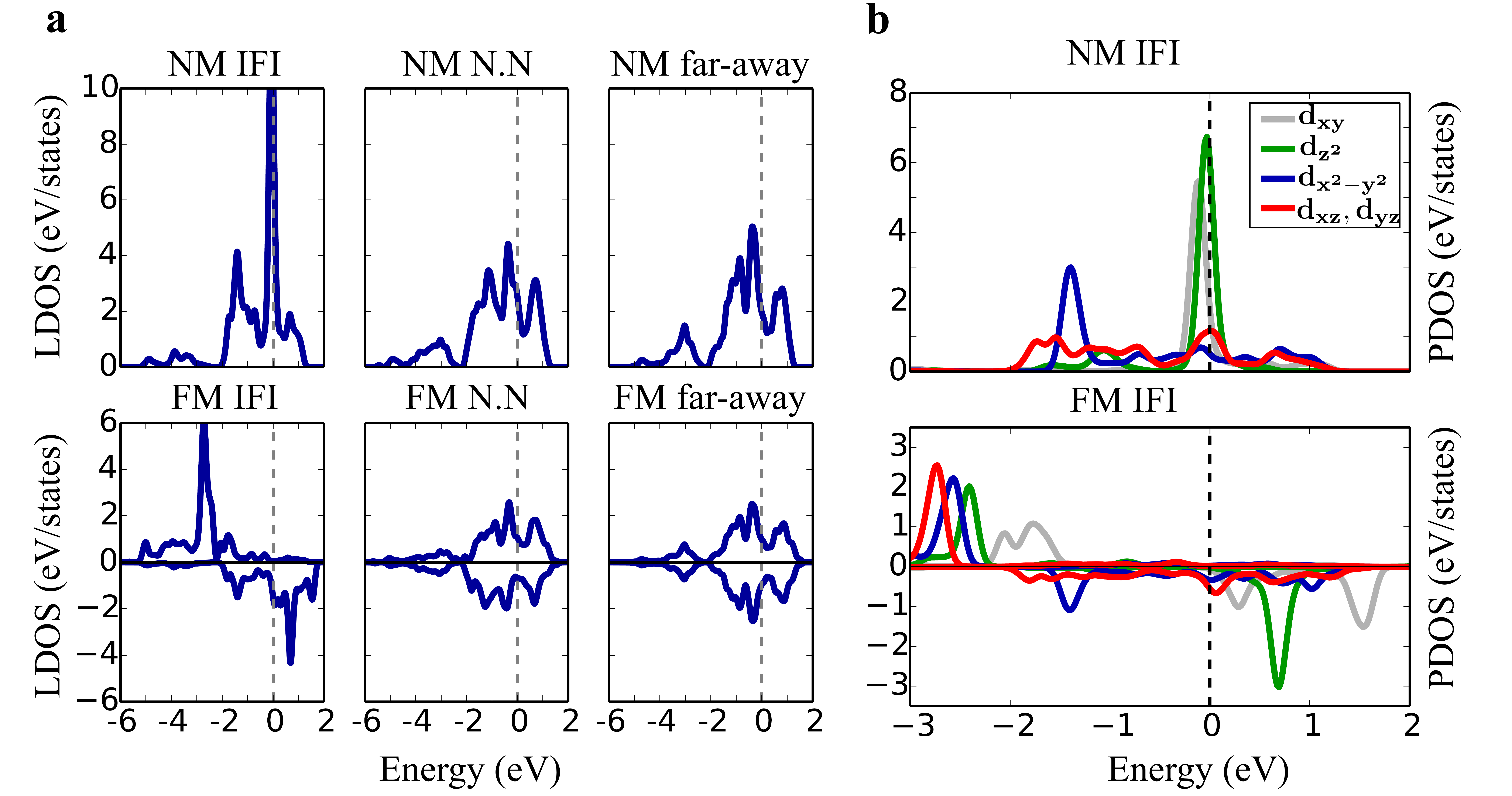}
\caption {
{\bf DFT-based local density of states.} {\bf a}, The local density of states on the IFI, one of its nearest-neighboring Fe sites, that far away into the pristine FeSe for a nonmagnetic (upper panel) and spin-polarized IFI (lower panel).  {\bf b}, The $d$-orbital-resolved partial density of states on the nonmagnetic (upper panel) and spin-polarized (lower panel) IFI site. The positive/negative value of each lower panel describes the up/down spin component.
}\label{figLDA}
\end{figure*}

\section{Local electronic structure around an adsorbed impurity}
The TSC property can also manifest in the local electronic structure around a single impurity.  The use of a single impurity is a powerful method to probe the superconducting pairing symmetry~\cite{AVBalatsky:2006}. More recently, a universal impurity-induced mid-gap state has also been predicted as a clear signature of topological superfluids~\cite{XJLiu:2012b,HHu:2013b}, where the near-zero-energy bound state follows closely  follow the symmetry of that of Majorana fermions. Inspired by the observation of a robust zero-bias conductance peak in the STM experiments around an  interstitial-Fe-impurity on the surface of the Fe(Te,Se) compound~\cite{SHPan}, we construct a microscopic model for the local adsorbed impurity within our  model. Figure~\ref{IFI}a shows the 2D lattice structure for a pristine system, where $d_{xz,yz}$-orbital orientation is considered; 
while Fig.~\ref{IFI}b  shows an adsorbed Fe atom and its effective local hopping with the neighboring Fe atoms in the parent compound.
We write down the total Hamiltonian with the local effect of absorbed site,
\begin{equation}
\mH^{Tot} = \mH^{TSC} +  \mT^{Imp}[\,\xi\,]+ \mS_z^{Imp}[\,J\,]\;,
\end{equation}
in which $\mT^{Imp}$ is the local hopping term from the impurity site to its nearest neighboring Fe atoms in the parent system with tunable hopping parameter $\xi$ (see Fig.~\ref{IFI}b), $\mS_z^{Imp}$ describes the local spin-polarized energy levels on the impurity site  with an exchange parameter $J$.
See more details for $\mT^{Imp}$ and $\mS^{Imp}_z$ in the {\bf Methods} Section.
We will focus on  the local density of states (LDOS) for fixed topological order $\lambda_{AOH} (=0.15 \text{ and } 0.0)$ and SC pairing strength $\Delta (=0.02)$ with varying $\xi$ and $J$.

In the regime of a weak hybridization, one can anticipate that the resulting peaks in the LDOS on the adsorbed impurity site merely reflects the energy levels of an isolated atom. In our work, we are more interested in the regime where the coupling between the adsorbed impurity and its neighbors is relatively strong.  Figure~\ref{figLDOS}a shows the LDOS calculated on the adsorbed impurity for various values of $\xi$ and $J$ but with fixed $\Delta=0.02$ and $\lambda_{AOH}=0.15$. For comparison, we also show the LDOS far away from the impurity site (black line in Fig.~\ref{figLDOS}a), which resembles the DOS in a pristine bulk.  For $J=0$, since the local hopping is still finite, two in-gap peaks appears near the gap edges. With the increased $J$, these in-gap peaks are shifted toward the Fermi energy (e.g.,  in the range of $J\in [ 0, 0.8]$ for $\xi=2.0$ as shown in Fig.~\ref{figLDOS}a), and cross the Fermi energy (e.g., at $J \sim 0.8$ for $\xi=2.0$ as shown in Fig.~\ref{figLDOS}a), and move toward the gap edges (e.g., in the range of $J\in [0.8, 2.0]$ for $\xi=2.0$ as shown in Fig.~\ref{figLDOS}a).  We also found a regime of $J \in [4,5]$, for which the in-gap quasiparticle peaks show up again.
Since in this large $J$ regime, the effect of the topological order becomes non-essential, our results are comparable with those in the early study on a magnetic impurity in an iron-based superconductor~\cite{WFTsai}. Therefore, we do not show here the results for this regime.
The in-gap bound states can also be visualized in the band structure in the reduced Brillouin zone of the supercell lattice~\cite{JXZhu1}.
In Fig.~\ref{figLDOS}b, all the sub-figures are plotted with varying $J$ for fixed values of $\xi=2$,  $\lambda_{AOH}=0.15$ and $\Delta=0.02$.
The localization nature of these in-gap states are investigated by different size of the supercell. As shown in Fig.~\ref{figLDOS}b, the dispersiveness of the in-gap states is suppressed with increased size of the supercell, suggesting that these in-gap states are truly bound states.  To further investigate the consequence of the topological order, we show in Fig.~\ref{figLDOS}c the LDOS on the impurity site in the absence ($\lambda_{AOH}=0$) and presence ($\lambda_{AOH}=0.15$) of the topological order.  The zero-energy peak in the LDOS in the absence of the topological order  is much narrower (accompanied with higher intensity) than that in the presence of the topological order, suggesting a much shorter decay length of quasiparticle bound state in the former.
This contrast is further confirmed by the spatial dependence of the LDOS at zero-energy, as shown in Fig.~\ref{figLDOS}d. Specifically, the zero-energy bound state has a longer decay length along the bond direction in a two-dimensional lattice. This difference can be understood from the band structure of the pristine system in an unfolded Brillouin zone, as shown in Fig.~\ref{figBdGBand}, where one can find that the topological order reduces significantly the gap for  the wave vector near the $X^{\prime\prime}$ point  along the $\Gamma-X^{\prime\prime}-M^{\prime\prime}$ path.

\section{Relevance to recent STM experiment}
We now discuss the relevance of our results to the recent STM experiment on Fe(Te,Se) compound for an adsorbed Fe impurity~\cite{SHPan}.
On the one hand,  the Fe(Te,Se) compound as mentioned in Ref.~\onlinecite{SHPan} has the number of $d$-electrons corresponding to the half-filling in the two-orbital model, which is different from the highly electron-doped KFe$_2$Se$_2$. Therefore, the chemical potential should be set to ensure the half-filling in our calculations.  In addition, since the distance from the adsorbed Fe site to its nearest neighbors are shorter than the Fe-Fe bonding distance of the parent compound, we anticipate that the local hopping parameter ($\xi$) between  the adsorbed Fe atom and its neighboring Fe atoms is larger than those for the parent compound. On the other hand, since the superconductivity for iron-based compounds arises from the Fe $3d$-electrons on an essentially square lattice, the normal-state part of the model Hamiltonian originally designed  for doped BaFe$_2$As$_2$~\cite{YTai1}, to match the low-energy band structure from the LDA calculations~\cite{ASubedi}, should also be applicable to the Fe(Te,Se) system. 

For the present purpose, we carried out the DFT calculations  for an Fe atom adsorbed  in the center of Fe plaquette of a large real-space structure of Fe(Te,Se) compound. See the \textbf{Methods} Section for the calculation details. We have found that a spin-polarized Fe atom adsorbed on the surface of an otherwise paramagnetic FeSe compound has a lower energy than a non-spin-polarized adsorbed Fe atom. 
Figure~\ref{figLDA}a shows the calculation of LDOS for both non-spin-polarized (upper panel) and spin-polarized (lower panel) adsorbed Fe atom. The broadened peak of the LDOS on the adsorbed Fe atom itself and the smearing of LDOS structure on its neighboring Fe sites in the FeSe compound (as compared with that on the Fe site far way from the adsorbed Fe atom)  suggests a strong coupling between the 
adsorbed Fe and its neighboring Fe sites.  It supports the treatment of the local hopping parameters in our model calculations.
For the case of the spin-polarized case, the magnetization is mostly localized on the adsorbed Fe itself with a proximately induced moment on its nearest neighboring Fe sites negligibly small. Finally, as shown in Fig.~\ref{figLDA}b, each $d$-orbital is spin polarized and contributes to the total magnetic moment of  2.89 $\mu$B. However,  among the total magnetic moment, only about 14.48\% (about $0.42 \;\mu_B$) arises from each of the $d_{xz}$- and $d_{yz}$-orbitals, which are relevant to the corresponding orbitals active in the pristine compound. These finding supports the use of  a local magnetic impurity with a small spin polarization in our model calculations.

\section{Discussion}
Our model study has shown the combined effect of the $Z_2$ topological mirror order and the fully gapped $s_{\pm}$-wave SC.
This $Z_2$ argument extends the mirror symmetry in the mirror-Chern representation~\cite{FZhang1} for the topological superconductivity of a 2D lattice.
The coexistence of this topological and the superconducting order relies on the intrinsic band structure of the $d_{xz,yz}$ orbitals, which gives  a topological metallic state. There the emergence of gapless edge states in the coexisting phase is a direct evidence for the topological order. In addition, the introduction of  a spin polarized impurity can be regarded as a zero-dimension  boundary to the 2D lattice, and it induces a similar quasiparticle behavior as around   a vortex core center in the topological superconductor~\cite{MSato}. 
In addition, our model calculations of the local electronic structure have revealed the topological feature of a superconductor through the LDOS peak intensity and the size of the  impurity-induced in-gap bound states. Finally, our DFT calculations of an adsorbed Fe atom on the surface of the FeSe compound has indeed shown the adsorbed impurity is spin polarized, making the findings in the present work relevant to  the STM experiment on the Fe(Te,Se) compound.

{\it Note added.} As we nearly completed our research, we noticed a recent study by Zhang~\cite{DZhang} that described the in-gap bound state around an adsorbed Fe atom on the surface of iron-based superconductors, within an Anderson impurity model with non-spin polarized $d$-electrons on the impurity site. Due to the mismatch of energy levels between the impurity and the Fe-sites in the background, the qualitative agreement with the STM experiment~\cite{SHPan} was obtained by adjusting the chemical potential in the pristine system in the strict condition of an unrealistically small hybridization between the adsorbed Fe atom and its neighbors. The shifted chemical potential leads to the system in the highly electron doped regime, which is inconsistent with the fact that Fe(Se,Te) is in the half-filled regime in the two-orbital model. In addition, it can be easily understood that with such a small hybridization, the zero-energy peak at the adsorbed Fe site has the character of the energy  levels of an isolated atom so that the peak intensity is too high. 

\section{Methods}

\begin{figure*}
\includegraphics[scale=0.5]{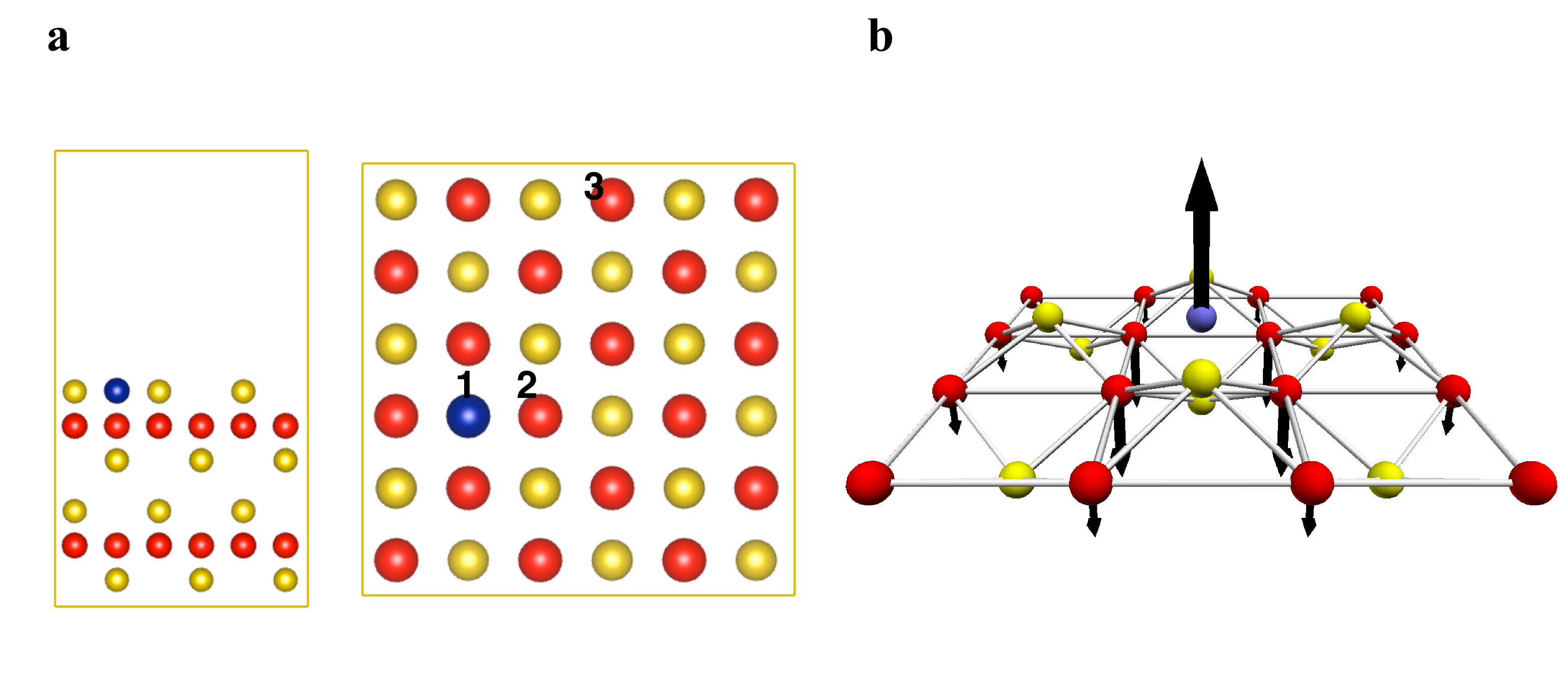}
\caption {
{\bf Setup of a two-layered with an adsorbed Fe impurity.}
{\bf a}, The side/top view of the real-space geometry. The labels 1/2/3 indicate the IFI, NN and far-away sites.
{\bf b}, Schematic picture showing the DFT calculated magnetic distribution (black arrow). The actual magnitude of each of them is: $2.89 \mu_B$ (IFI site), $0.23 \mu_B$ (NN site) and $0.019 \mu_B$ (NNN site).
In each panel the red/yellow balls represent Fe/Se atoms while the blue one stands for the IFI.
}\label{VASP}
\end{figure*}

\textbf{Model Setup.~} Here we write down the details of $\mH^{TS}=\mT[t_{1-6},\mu]+\mV[\lambda_{AOH}]+\mP[\Delta]$,
\begin{equation}
\begin{aligned}
\mT&=\sum_{IJ,\alpha\beta,s} c^\dagger_{I\alpha s}\, (t^{\alpha\beta}_{IJ}-\mu\;\delta_{IJ}\delta_{\alpha\beta})\, c_{J\beta s},\\
\mV&=\sum_{\langle IJ\rangle,\alpha,s} i(-1)^{s}\lambda_{AOH}\;\nu^{\alpha\bar\alpha}_{IJ}\; c^\dagger_{I\alpha s}\, c_{J\bar\alpha s},\\
\mP&=\sum_{IJ,\alpha,s} (\Delta\,c^\dagger_{I\alpha s}\, \, c^\dagger_{J\alpha s'}+H.c),
\end{aligned}
\end{equation}
were $I,J$ are site index, $\alpha,\beta\in [d_{xz}, d_{yz}]$ are orbital index, $s\in [\uparrow, \downarrow]$ is the spin index and $\mu$ is the chemical potential to adjust the Fermi surface for hall-filling ($n=2$).
Here, $t^{\alpha\beta}_{IJ}\in t_{1-6}$, are the hopping integrals~\cite{YTai1,HChen}.
In our previous paper~\cite{YTai2} $\lambda_{AOH}$ could be obtained self-consistently from the NN inter-orbital Coulomb interaction at the mean-field level. However, in this work, we follow the same spirit as ref.~\onlinecite{JXZhu} and treat $\lambda_{AOH}$ and superconducting pair potential $\Delta$ as input parameters for our analysis.
The tensor elements $\nu_{IJ}^{\alpha\bar\alpha} \in [0,\pm 1]$ describe the direction of the NN inter-orbital currents with $\nu_{\pm\hat x}^{12}=\nu_{\pm\hat y}^{21}=-1$, and $\nu_{\pm\hat x}^{21}=\nu_{\pm\hat y}^{12}=1$.

The impurity model, $\mH^{Imp}=\mT^{Imp}[\xi]+\mS^{Imp}_z[J]$, is written,
\begin{equation}
\begin{aligned}
	\mT^{IFI}	&= \sum_{\delta,\alpha,s} t_{\delta\alpha} (c^\dagger_{o,\alpha,s}\, c_{\delta,\alpha,s}+H.c),\\
	\mS^{IFI}_z	&= (J/2) \sum_{\alpha,ss'} c^\dagger_{o,\alpha,s} (\hat z\cdot \vec \sigma_{ss'}) c_{o,\alpha,s'},\\
				&= (J/2) \sum_{\alpha} (n_{o,\alpha,\uparrow}-n_{o,\alpha,\downarrow}).
\end{aligned}
\end{equation}
The index, $o$, denotes the impurity site and $\delta$ indicates the NN sites from $o$ to its NN sites.
$t_{\delta\alpha}$ is the local hopping integrals from the IFI site to its NN sites.
Based on the d$_{xz,yz}$ orbital orientation, as shown in Fig.~\ref{IFI}b, there are only two different values in our impurity model $t_{\delta\alpha}\in t_\sigma, t_\pi$; here we take $t_\sigma=a\,t_\pi=\xi$ with $a=-3$.
In our calculation, different sign and magnitude of $a$ do not change our results qualitatively.

The LDOS can be calculated according to,
\begin{equation}
\rho_{i\alpha}(E)=\frac{1}{M}\sum_{n,{\bf k}} [|u_{i\alpha}^{n,{\bf k}}|^2 \delta(E_{n,{\bf k}}-E)+|v_{i\alpha}^{n,{\bf k}}|^2\delta(E_{n,{\bf k}}+E)],
\end{equation}
where the $u^{n,{\bf k}}_{i\alpha}$ and $v^{n,{\bf k}}_{i\alpha}$ are the eigenfunctions of the BdG matrix of the entire Hamiltonian.
We use the broadening facor $\Gamma=0.001$ in $\delta(x)=\Gamma/\pi(x^2+\Gamma^2)$.
The supercell technique~\cite{JXZhu1} is used for $M$-repeated cell blocks.

%

\textbf{DFT Calculations.~} 
We use the Vienna Ab-initio Simulation Package (VASP)~\citep{vasp} to carry out the LDA calculations for an interstitial Fe impurity (IFI) atom sitting in the center of a large real-space structure of Fe(Te,Se) compound.  
The projector augmented planewave method~\cite{paw} and the Perdew-Burke-Ernzerhof~\cite{pbe} exchange-correlation functional were adopted.
We used  a 4$\times$4$\times$2 Monkhorst-Pack k-point mesh, and a 500 eV cutoff. 
The $z$-coordinate of Fe impurity atoms was fully relaxed  up to $10^{-2}$ eV/$\AA$ with other atoms fixed.

We have considered the setup as shown in Fig.~\ref{VASP}a, in which an IFI atom sitting on the top of a two-layered FeSe. 
A vacancy space in the z-direction was made to mimic the surface effect of the single IFI atom.
We have calculated({\bf i}) non-magnetic(NM) state and ({\bf ii}) ferro-magnetic(FM) state on the IFI site by turning on and off spin polarization initially on IFI site. Each of the two calculations has reached the charge convergence. 
Fig.~\ref{VASP}b shows the local distribution of magnetic moment.
The calculated total energy shows that the case ({\bf ii}) is more stable than the case ({\bf i}), which confirms a spin polarized IFI site.

\section{Acknowledgements}
We thank S.H.Pan for useful discussion in an early stage of our research.
The work at Los Alamos was supported by the U.S.\ DOE Contract No.~DE-AC52-06NA25396 through the
LDRD Program (Y.-Y.T. \& H.C.), the Office of Basic Energy Sciences (BES) (T.A. \& J.-X.Z.).
This work was supported in part by the Center for Integrated Nanotechnologies, a DOE BES user facility (J.-X.Z.).
The work at University of Houston was supported in part by the Robert A.\ Welch Foundation under Grant No.\ E-1146 and AFOSR under Grant No. FA9550-09-1-0656 (C.S.T.).

\end{document}